\documentclass[showpacs,prb,twocolumn,superscriptaddress]{revtex4}

\usepackage{amsmath}
\usepackage{graphicx}
\usepackage{dcolumn}
\usepackage{bm}

\begin{document}
\title{Two-spin entanglement induced by electron scattering in nanostructures}

\author{Gian Luca Giorgi}\email{gianluca.giorgi@roma1.infn.it}
\affiliation{CNR-INFM Center for Statistical Mechanics and
Complexity} \affiliation{Dipartimento di Fisica, Universit\`{a} di
Roma La Sapienza, Piazzale A. Moro 2, 00185 Roma, Italy}
\author{Ferdinando  de Pasquale}
\affiliation{Dipartimento di Fisica, Universit\`{a} di Roma La
Sapienza, Piazzale A. Moro 2, 00185 Roma, Italy}
\affiliation{CNR-INFM Center for Statistical Mechanics and
Complexity}


\begin{abstract}
We present a model where two magnetic impurities in a discrete
tight-binding ring become entangled because of scattering processes
associated to the injection of a conduction electron. We introduce a
weak-coupling approximation that allows us to solve the problem in a
analytical way and compare the theory with the exact numerical
results. We obtain the generation of entanglement both in a
deterministic way and in a probabilistic one. The first case is
intrinsically related to the structure of the two-impurity reduced
density matrix, while the second one occurs when a projection on the
electron state is performed.

\end{abstract}
\maketitle

\section{Introduction}

The generation of entanglement in mesoscopic structures is
considered as a fundamental resource for the implementation of
solid-state quantum information processing devices \cite{nielsen}.
The first proposals for spin-based quantum computation concern
direct interaction between qubits \cite {loss}. Further, different schemes for
mesoscopic structures have been suggested that create
separated streams of entangled particles \cite
{lebovitz,bena,sukho,burkard,recher,samuelsson}.

Recently, Costa {\it et al.} \cite{costa} have examined the
possibility of entangling two spatially separated stationary spins
by means of electron scattering. In this case the generation of
entanglement would require lower external control. After considering
a toy model concerning a ballistic electron interacting in
succession with two distant spins, Costa {\it et al.} face the more
realistic problem of how two magnetic impurities embedded in a solid
become entangled because of the injection of a conduction electron,
which is scattered by the impurities according to a {\it s-d}
Hamiltonian. Since in a normal metal described by a tight-binding
model the energy spectrum constitutes a continuous band, the
appearance of an imaginary part in the eigenenergies of the system
is expected that would limit the coherence time of the entangled
state. A way to overcome this instability can be represented by the
introduction of an artificial discrete system, i.e. a ring of
$N$-coupled quantum dots (QDs). During the last decade, electronic transport
properties through quantum dots have been widely considered both
experimentally and theoretically \cite{dot}. Because of the progress
of nanotechnology, it is possible to fabricate various structures of
coupled QDs  smaller than the electron coherence length.

If the interaction term between the electron spin and the impurities is much
less than the energy separation between consecutive eigenvalues, in a
finite-size system such as a nanostructure, resonance conditions are reached, and a
reduction to a few-body system \cite{noi} can be observed. In that case an
oscillatory regime is expected to come out also in the degree of
entanglement. Dissipation effects could appear only through the interaction
with some external bath (for instance, coupling with phonons).

The aim of this paper is to show how entanglement through electron
scattering can be generated efficiently in such nanostructures.
Then, we consider a finite tight-binding model, and add two magnetic
impurities. By studying the dynamical evolution of the state of a
conduction electron injected in the chain, we establish the amount
of entanglement between the two spins as a function of time. We find
that entanglement can be generated in a deterministic way as well as
in a probabilistic one.

The plan of the work is the following. After the Introduction, in
Sec. \ref{II} we define the general Hamiltonian model, getting the
equations of motion for the states involved in the evolution. In
Sec. \ref{III} we establish the approximation of considering the
interaction of the incoming mode only with the other resonant modes,
and solve the equation of motion. In Sec. \ref{IV} the generation of
entanglement is studied by means of two different approaches: the
measure of concurrence associated to the reduced density matrix
obtained by tracing out the electron spin, and the ``localizable
entanglement" derived from a projective measurement on the
electron. In Sec. \ref{V} we conclude the paper.

\section{Model} \label{II}
\begin{figure}
\begin{center}
  \includegraphics[width=9cm]{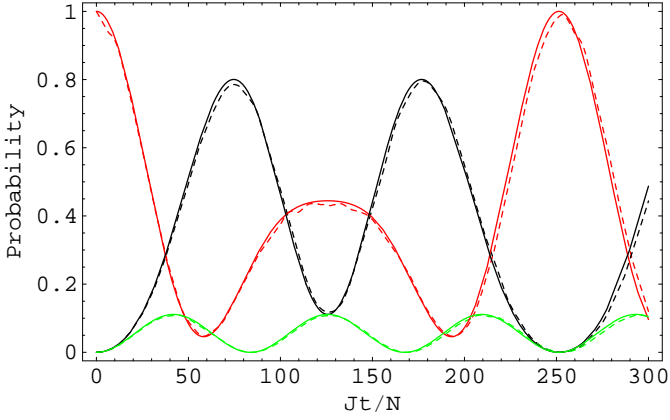}\\
  \caption{Plot of the probabilities of the states involved
  in the evolution as a function of time. Solid lines correspond to theoretical predictions,
   while dashed lines show numerical (exact) evolution.
   Red lines concern the state $\left| \uparrow \uparrow
\right\rangle \left| \downarrow _{k}\right\rangle $, black lines are
related to $\left| \uparrow \uparrow \right\rangle \left| \downarrow
_{-k}\right\rangle $, while green lines regard $\left| \downarrow
\uparrow \right\rangle \left| \uparrow _{k}\right\rangle $. The
probabilities associated to $\left| \downarrow \uparrow
\right\rangle \left| \uparrow _{-k}\right\rangle $, $\left| \uparrow
\downarrow \right\rangle \left| \uparrow _{k}\right\rangle $, and
$\left| \uparrow \downarrow \right\rangle \left| \uparrow
_{-k}\right\rangle $ are not plotted, being practically
indistinguishable from that of $\left| \downarrow \uparrow
\right\rangle \left| \uparrow _{k}\right\rangle $. The system
parameters are the following: $w = 1$ is chosen as unit of energy;
the number of sites is $N = 16$;
  the scattering amplitude is $J=0.2$, while the distance
  between the spins is $L=4$.}\label{evolution}
  \end{center}
\end{figure}
We consider a discrete ring structure, described by a standard
tight-binding Hamiltonian, where two particular sites (for instance,
we label the first site with $0$ and the second one with $L$) are
substituted by magnetic impurities. The {\it s-d} Hamiltonian
describing the system is
\begin{equation}
H=H_{0}+\frac{J}{2}\left( \vec{S}_{0}\cdot \vec{\sigma}_{0}+\vec{S}_{L}\cdot
\vec{\sigma}_{L}\right) ,
\end{equation}
with $H_{0}=\sum_{k,\sigma }\epsilon _{k}a_{k,\sigma }^{\dagger
}a_{k,\sigma }$, where $a_{k,\sigma }^{\dagger }$ ($a_{k,\sigma }$)
creates (annihilates) one electron with spin $\sigma=\uparrow,\downarrow $ on the mode
$k$, $\epsilon _{k}=-2w\cos k$ are the eigenvalues of $H$ in the absence
of spin interaction ($w$ is the hopping amplitude between adjacent
sites, $k=\left( 2\pi /N\right) n$, $N$
is the total number of sites, and $n$ is an integer running from $-N/2$ to $%
[\left( N/2\right) -1]$), $J$ is the coupling constant between the
impurity
spins $\vec{S}_{0}$ and $\vec{S}_{L}$ and the electron spins $\vec{\sigma}%
_{0}$ and $\vec{\sigma}_{L}$, whose operators are defined as
\begin{eqnarray}
\sigma _{l}^{x} &=&a_{l,\uparrow }^{\dagger }a_{l,\downarrow
}+a_{l,\downarrow }^{\dagger }a_{l,\uparrow }  \nonumber \\
&=&\frac{1}{N}\sum_{q,q^{\prime }}\left( a_{q,\uparrow }^{\dagger
}a_{q^{\prime },\downarrow }+a_{q,\downarrow }^{\dagger }a_{q^{\prime
},\uparrow }\right) e^{i\left( q-q^{\prime }\right) l}, \\
\sigma _{l}^{y} &=&-i\left( a_{l,\uparrow }^{\dagger }a_{l,\downarrow
}-a_{l,\downarrow }^{\dagger }a_{l,\uparrow }\right)   \nonumber \\
&=&\frac{-i}{N}\sum_{q,q^{\prime }}\left( a_{q,\uparrow }^{\dagger
}a_{q^{\prime },\downarrow }-a_{q,\downarrow }^{\dagger }a_{q^{\prime
},\uparrow }\right) e^{i\left( q-q^{\prime }\right) l}, \\
\sigma _{l}^{z} &=&a_{l,\uparrow }^{\dagger }a_{l,\uparrow }-a_{l,\downarrow
}^{\dagger }a_{l,\downarrow }  \nonumber \\
&=&\frac{1}{N}\sum_{q,q^{\prime }}\left( a_{q,\uparrow }^{\dagger
}a_{q^{\prime },\uparrow }-a_{q,\downarrow }^{\dagger }a_{q^{\prime
},\downarrow }\right) e^{i\left( q-q^{\prime }\right) l}.
\end{eqnarray}
Let us consider the impurities initially with spin up,\ and the
introduction of one excess electron in the state $\left| \downarrow
_{k}\right\rangle $. Then, differently from the treatment given in
Ref. \onlinecite{costa}, we look at entanglement generation from a
dynamical point of view, i.e., we analyze the evolution in the time
domain of the state $\left| \uparrow \uparrow \right\rangle \left|
\downarrow _{k}\right\rangle $ to show explicitly how coherent
effects persist. By introducing the complex Laplace transform the
state evolves in
\begin{eqnarray}
\left| \uparrow \uparrow \right\rangle \left| \downarrow _{k}\right\rangle
_{\omega } &=&\frac{1}{\omega -\epsilon _{k}}[\left| \uparrow \uparrow
\right\rangle \left| \downarrow _{k}\right\rangle +  \nonumber \\
&&\frac{J}{N}\sum_{q}\left( \left| \downarrow \uparrow \right\rangle \left|
\uparrow _{q}\right\rangle _{\omega }+e^{i\left( q-k\right) L}\left|
\uparrow \downarrow \right\rangle \left| \uparrow _{q}\right\rangle _{\omega
}\right)   \nonumber \\
&&-\frac{J}{2N}\sum_{q}\left( 1+e^{i\left( q-k\right) L}\right) \left|
\uparrow \uparrow \right\rangle \left| \downarrow _{q}\right\rangle _{\omega
}]\label{a},
\end{eqnarray}
where the notation $\left| \right\rangle $ refers to a
configuration, while $\left| \right\rangle _{\omega }$ refers to
the evolution of that configuration, and $\hbar=1$. The other states involved in
the evolution satisfy the following equations:
\begin{eqnarray}
\left| \downarrow \uparrow \right\rangle \left| \uparrow _{k}\right\rangle
_{\omega } &=&\frac{1}{\omega -\epsilon _{k}}[\left| \downarrow \uparrow
\right\rangle \left| \uparrow _{k}\right\rangle   \nonumber \\
&&+\frac{J}{2N}\sum_{q}\left( -1+e^{i\left( q-k\right) L}\right) \left|
\downarrow \uparrow \right\rangle \left| \uparrow _{q}\right\rangle _{\omega
}  \nonumber \\
&&+\frac{J}{N}\sum_{q}\left| \uparrow \uparrow \right\rangle \left|
\downarrow _{q}\right\rangle _{\omega }]\label{b}, \\
\left| \uparrow \downarrow \right\rangle \left| \uparrow _{k}\right\rangle
_{\omega } &=&\frac{1}{\omega -\epsilon _{k}}[\left| \uparrow \downarrow
\right\rangle \left| \uparrow _{k}\right\rangle +  \nonumber  \label{c} \\
&&\frac{J}{2N}\sum_{q}\left( 1-e^{i\left( q-k\right) L}\right) \left|
\downarrow \uparrow \right\rangle \left| \uparrow _{q}\right\rangle _{\omega
}  \nonumber \\
&&+\frac{J}{N}\sum_{q}e^{i\left( q-k\right) L}\left| \uparrow \uparrow
\right\rangle \left| \downarrow _{q}\right\rangle _{\omega }]\label{c}.
\end{eqnarray}
\section{Resonant coupling approximation}\label{III}

The problem is significantly simplified by introducing the following
weak-coupling approximation. In the solution of the system derived
from Eqs. (\ref{a})-(\ref{c}), it would appear denominators
with the structure $\left[ \omega -\epsilon _{k}-\left(
J^{2}/N^{2}\right) \sum_{q}f\left(
k-q\right) /\left( \omega -\epsilon _{q}\right) \right] ^{-1}$, where $%
f\left( k-q\right) $ is some weight function derived from $\left(
\pm 1\pm e^{i\left( q-k\right) L}\right) $. If the scattering
amplitude, of the order of $J/N$, is much less than the energy
differences appearing in the spectrum of $H_{0}$, which is about
$w/N$ near the middle of the band, we keep just the resonant terms,
corresponding to $q=\pm k$ \cite{noi}. For instance,
\begin{eqnarray}
\frac{1}{\omega -\epsilon
_{k}-\frac{J^{2}}{N^{2}}\sum_{q}\frac{\left(
1-e^{i\left( q-k\right) L}\right) \left( 1+e^{i\left( q-k\right) L}\right) }{%
\omega -\epsilon _{q}}}\nonumber\\\simeq \frac{1}{\omega -\epsilon _{k}-\frac{J^{2}}{N^{2}}\frac{%
1-e^{2i\left( q-k\right) L}}{\omega -\epsilon _{k}}}
\nonumber\\=\frac{\omega -\epsilon _{k}}{\left( \omega -\epsilon _{k}\right) ^{2}-%
\frac{J^{2}}{N^{2}}\left[ 1-e^{2i\left( q-k\right) L}\right] },
\end{eqnarray}
and in the right-hand side of Eqs. (\ref{a})-(\ref{c}) we
maintain only states with momentum $k$ or $-k$, that is, the states
that were degenerate with the initial one in absence of
interaction. Under these assumptions the equations of motion reduce
to
\begin{widetext}
\begin{eqnarray}
\left| \uparrow \uparrow \right\rangle \left| \downarrow
_{k}\right\rangle _{\omega } &=&\frac{1}{\omega -\epsilon
_{k}+\frac{J}{N}}\left[ \left|
\uparrow \uparrow \right\rangle \left| \downarrow _{k}\right\rangle -\frac{J%
}{2N}\left( 1+e^{-2ikL}\right) \left| \uparrow \uparrow
\right\rangle \left|
\downarrow _{-k}\right\rangle _{\omega }\right]   \nonumber \\
&&+\frac{1}{\omega -\epsilon _{k}+\frac{J}{N}}\frac{J}{N}\left[
\left| \downarrow \uparrow \right\rangle \left( \left| \uparrow
_{k}\right\rangle _{\omega }+\left| \uparrow _{-k}\right\rangle
_{\omega }\right) +\left| \uparrow \downarrow \right\rangle \left(
\left| \uparrow _{k}\right\rangle
_{\omega }+e^{-2ikL}\left| \uparrow _{-k}\right\rangle _{\omega }\right) %
\right],  \\
\left| \downarrow \uparrow \right\rangle \left| \uparrow
_{k}\right\rangle _{\omega } &=&\frac{1}{\omega -\epsilon
_{k}}\left[ \left| \downarrow \uparrow \right\rangle \left| \uparrow
_{k}\right\rangle +\frac{J}{2N}\left( -1+e^{-2ikL}\right) \left|
\downarrow \uparrow \right\rangle \left| \uparrow _{-k}\right\rangle
_{\omega }+\frac{J}{N}\left| \uparrow \uparrow \right\rangle \left(
\left| \downarrow _{k}\right\rangle _{\omega }+\left|
\downarrow _{-k}\right\rangle _{\omega }\right) \right] , \\
\left| \uparrow \downarrow \right\rangle \left| \uparrow
_{k}\right\rangle _{\omega } &=&\frac{1}{\omega -\epsilon
_{k}}\left[ \left| \uparrow
\downarrow \right\rangle \left| \uparrow _{k}\right\rangle +\frac{J}{2N}%
\left( 1-e^{-2ikL}\right) \left| \uparrow \downarrow \right\rangle
\left| \uparrow _{-k}\right\rangle _{\omega }+\frac{J}{N}\left|
\uparrow \uparrow \right\rangle \left( \left| \downarrow
_{k}\right\rangle _{\omega }+e^{-2ikL}\left| \downarrow
_{-k}\right\rangle _{\omega }\right) \right] .\nonumber\\
\end{eqnarray}
\end{widetext}
Together with these equations we must consider also those obtained
by exchanging $k$ with $-k$.

A further simplification can be introduced by properly choosing the
distance between impurities $L$. For instance, if $k=\pi /2$ and $L$
is even, $e^{\pm 2ikL}=1$. In this case the study of the state
evolution greatly simplifies. Since at $t=0$ we had $\left| \uparrow
\uparrow \right\rangle \left| \downarrow _{k}\right\rangle $, at the
time $t$ we obtain

\begin{widetext}
\begin{eqnarray}
\left| \uparrow \uparrow \right\rangle \left| \downarrow
_{k}\right\rangle _{t} &=&\frac{1}{6}\left(
3+e^{-2i(J/N)t}+2e^{4i(J/N)t}\right)
\left| \uparrow \uparrow \right\rangle \left| \downarrow _{k}\right\rangle \nonumber \\&&+%
\frac{1}{6}\left( -3+e^{-2i(J/N)t}+2e^{4i(J/N)t}\right)
\left| \uparrow \uparrow \right\rangle \left| \downarrow
_{-k}\right\rangle
\nonumber \\
&&+\frac{1}{6}e^{-2i(J/N)t}\left( 1-e^{6i(J/N)t}\right)
\left( \left| \downarrow \uparrow \right\rangle +\left| \uparrow
\downarrow \right\rangle \right) \left( \left| \uparrow
_{k}\right\rangle +\left| \uparrow _{-k}\right\rangle \right).
\label{evolu}
\end{eqnarray}
\end{widetext}
The correctness of the above approximation is checked by numerical
integration of the Hamiltonian evolution. In Fig. \ref{evolution} we
compare the probabilities derived from the coefficients in Eq.
(\ref{evolu}) with the exact results. The agreement between
perturbation theory and numerical results is remarkable.

\section{Entanglement measures}\label{IV}

The two-impurity reduced density matrix $\rho $ is obtained by
tracing out the electron degree of freedom: in the basis spanned by
the states $\{\left| \uparrow \uparrow \right\rangle ,\left|
\downarrow \uparrow \right\rangle ,\left| \uparrow \downarrow
\right\rangle ,\left| \downarrow \downarrow \right\rangle \}$ we
have
\begin{equation}
\rho =\left(
\begin{array}{cccc}
1-\frac{4}{9}\sin ^{2}\frac{3J}{N}t & 0 & 0 & 0 \\
0 & \frac{2}{9}\sin ^{2}\frac{3J}{N}t & \frac{2}{9}\sin
^{2}\frac{3J}{N}t & 0
\\
0 & \frac{2}{9}\sin ^{2}\frac{3J}{N}t & \frac{2}{9}\sin ^{2}\frac{3J}{N}t & 0
\\
0 & 0 & 0 & 0
\end{array}
\right).
\end{equation}

\begin{figure}[tb]
\begin{center}
  \includegraphics[width=9cm]{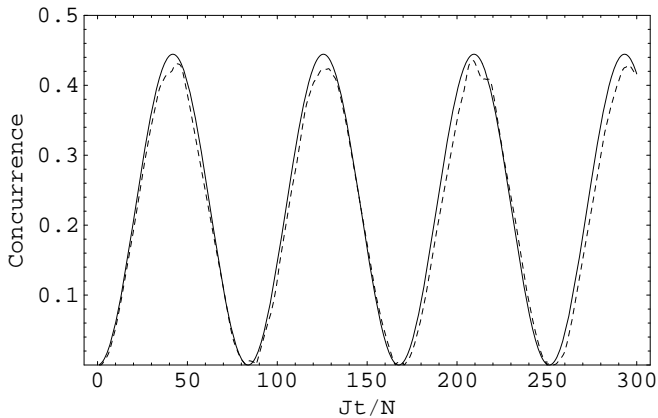}\\
  \caption{Concurrence as a function of time. The solid line is derived from the theoretical
  model described in the text, while the dashed line corresponds to the numerical
  calculation. The system parameters are the same as those defined
   in the caption of Fig. \ref{evolution}.}\label{concurrence}
  \end{center}
\end{figure}

Given $\rho $, we can compute the corresponding degree of
entanglement by means of the concurrence ${\cal C}$ \cite{wootters}.
The concurrence between two qubits is defined to be ${\cal C}=\max
\{\lambda _{1}-\lambda _{2}-\lambda _{3}-\lambda _{4},0\}$, where
$\lambda _{r}$ is the square root of the {\it r}th eigenvalue of $R=\rho
\tilde{\rho}$ in descending order. The matrix $\tilde{\rho}$ is
defined as  $\tilde{\rho}=(\sigma _{y}\otimes \sigma _{y})\rho
^{\ast }(\sigma _{y}\otimes \sigma _{y})$, where $\rho ^{\ast }$ is
the complex conjugate of $\rho $. Since the eignevalues of $R$ are
$\left\{ 0,0,0,\left( 16/81\right) \sin ^{4}\left( 3Jt/N\right)
\right\} $, the corresponding $\lambda _{r}$ are $\left\{
0,0,0,\left( 4/9\right) \sin ^{2}\left( 3Jt/N\right) \right\} $. So
we have ${\cal C}=4/9\sin ^{2}\left( 3Jt/N\right) $. In Fig.
\ref{concurrence} we report the numerical value of ${\cal C}$\ as a
function of time and compare it with the analytical expression.
In this way we have calculated the amount of entanglement arising
spontaneously from scattering processes. Actually, there is a
different strategy from which a higher degree of entanglement, the
so-called localizable entanglement \cite{cirac,cirac2}, could be
extracted. It consists in a projective measurement performed on the
electron degree of freedom. If we project $\left| \uparrow \uparrow \right\rangle \left| \downarrow
_{k}\right\rangle _{t} $ onto
$\left( \left| \uparrow _{k}\right\rangle +\left| \uparrow
_{-k}\right\rangle \right) /\sqrt{2}$ we get the state
\begin{equation}
\frac{\left\langle \uparrow _{k}\right| + \left\langle \uparrow
_{-k}\right|}{\sqrt{2}}\left| \uparrow \uparrow \right\rangle
\left| \downarrow _{k}\right\rangle _{t}=\frac{ e^{-2i(J/N)%
t}-e^{4i(J/N)t}}{3} \frac{\left| \downarrow \uparrow
\right\rangle +\left| \uparrow \downarrow \right\rangle }{\sqrt{2}}.
\label{proj}
\end{equation}
As a result of the projective measurement, a maximally entangled
state appears for all the times. However, the probability $P$\ of
actually finding $\left( \left| \uparrow _{k}\right\rangle +\left|
\uparrow _{-k}\right\rangle \right) /\sqrt{2}$ is different from
one;
\begin{equation}
P=\left| \frac{1}{3}\left(
e^{-2i(J/N)t}-e^{4i(J/N)t}\right) \right|
^{2}=\frac{4}{9}\sin ^{2}\frac{3J}{N}t.
\end{equation}
That is, the process is probabilistic instead of deterministic. We
obtain a success probability that evolves in time with the same law
of ${\cal C}$. The two results have the following interpretation. In
the first case the two spins are spontaneously entangled by electron
scattering. The amount of entanglement is obviously related to the
probability of finding the component $\left( \left| \downarrow
\uparrow \right\rangle \left| \uparrow _{k}\right\rangle +\left|
\downarrow \uparrow \right\rangle \left| \uparrow _{-k}\right\rangle
+\left| \uparrow \downarrow \right\rangle \left| \uparrow
_{k}\right\rangle +\left| \uparrow \downarrow \right\rangle \left|
\uparrow _{-k}\right\rangle \right) $. In the physical procedure
related to the projective measurement, we consider only this
component. The same time evolution is then associated to two
different kinds of processes, the first one being deterministic and
the second one being probabilistic. In fact, without any kind of
projection, we have a limited degree of entanglement. By projecting
on the electron state we can reach ${\cal C}=1$, but the price to
pay consists in a limited success probability of the projection. The
best strategy to adopt will depend on the specific application.

\section{Conclusions}\label{V}

In this paper we have discussed the problem of entangling two
distant spins embedded in a solid-state environment through the
interaction with a conduction electron. We have described explicitly
what happens when a finite-size system is considered. By applying a
weak-coupling approximation, that is, by neglecting nonresonant
scattering states, we have solved analytically the evolution in time
of the state associated to an incoming conduction electron. In fact,
the existence of discrete levels and the weakness of the coupling
make possible a resonance between the scattered states and one level
of the energy band. A comparison between theoretical results and
exact numerical results has been presented, showing the accuracy of
the weak-coupling approximation. As a result of the evolution,
two-spin entanglement appears. We have analyzed the emergence of
entanglement merely from evolution through the reduced density
matrix, and the role of projection on the electron spin state. In
conclusion, magnetic scattering in a discrete system has been shown
to create entanglement both in a deterministic and in a
probabilistic way.

\end{document}